\documentclass[journal,twocolumn]{IEEEtran}
\usepackage{epsfig,makeidx,color}
\usepackage{amsmath,amssymb,amsthm,bbm}
\usepackage{cite,graphicx}
\usepackage{enumerate}
\usepackage{hyperref}
\hypersetup{
        colorlinks = true,
        citecolor=blue,
}

\def\cK{{\cal K}}

\def\cZ{{\cal Z}}

\def\rT{{\rm T}}

\def\uE{{\mathbb E}}

\DeclareMathOperator*{\argmin}{\arg\!\min}
\DeclareMathOperator*{\argmax}{\arg\!\max}

\def\be{ \begin{equation} }
\def\ee{ \end{equation} }
\def\bea{ \begin{eqnarray} }
\def\eea{ \end{eqnarray} }
\def\bx{{\bf x}}

\def\bc{{\bf c}}

\def\bb{{\bf b}}

\def\ba{{\bf a}}

\def\bm{{\bf m}}

\def\bz{{\bf z}}

\def\bI{{\bf I}}

\def\bR{{\bf R}}

\def\b0{{\bf 0}}

\def\cK{{\cal K}}

\def\cN{{\cal N}}

\def\sD{{\sf D}}

\def\sH{{\sf H}}

\def\sSNR{{\sf SNR}}

\ifCLASSOPTIONonecolumn
  \newcommand{\figwidth}{0.40\columnwidth}
\else
  \newcommand{\figwidth}{0.80\columnwidth}
\fi

\begin{document}

\title{Data-aided Sensing for Distributed Detection}

\author{Jinho Choi
\thanks{The author is with
the School of Information Technology,
Deakin University, Geelong, VIC 3220, Australia
(e-mail: jinho.choi@deakin.edu.au).
This research was supported
by the Australian Government through the Australian Research
Council's Discovery Projects funding scheme (DP200100391).}}

\maketitle

\begin{abstract}
In this paper, we study data-aided sensing
(DAS) for distributed detection in wireless sensor networks (WSNs)
when sensors' measurements are correlated.
In particular, we derive a node selection criterion based on the J-divergence
in DAS for reliable decision subject to a decision delay constraint.
Based on the proposed J-divergence based DAS,
the nodes can be selected to rapidly increase
the log-likelihood ratio (LLR), which leads to a reliable decision 
with a smaller number of the sensors that upload measurements
for a shorter decision delay.
From simulation results,
it is confirmed that 
the J-divergence based DAS can provide
a reliable decision 
with a smaller number of sensors compared to other approaches.
\end{abstract}

{\IEEEkeywords
Distributed Detection;
Wireless Sensor Networks;
Intelligent Data Collection}

\ifCLASSOPTIONonecolumn
\baselineskip 24pt
\fi

\section{Introduction} \label{S:Intro}

In wireless sensor networks (WSNs), a set of sensors 
are deployed to perform environmental monitoring and send
observations or measurements to a fusion center (FC)
\cite{Dargie10}.
The FC is responsible for making a final decision 
on the physical phenomenon from the reported information.
WSNs can be part of the Internet of Things (IoT) \cite{Zhu10}
\cite{Kuo18} to provide data sets collected from 
sensors to various IoT applications.

In conventional WSNs, various approaches are studied
to perform distributed detection 
with decision fusion rules at the FC
\cite{Varshney97Distributed}.
The optimal decision fusion rules in various 
channel models have been intensively studied 
in the literature (see references in \cite{Visw97}
\cite{Chamberland03Decentralized}).
In \cite{Chen06}, efficient transmissions
from sensors to the FC 
studied using the channel state information (CSI).
Furthermore, in \cite{He10},
cross-layer optimization is considered 
to maximize the life time of WSN
when sensors' measurements are correlated.
In \cite{Chawla19},
the notion of massive multiple input multiple output
(MIMO) \cite{Marzetta10} is exploited for distributed
detection when the FC is equipped with a large
number of antennas.

In a number of IoT applications, as in WSNs,
devices' sensing to acquire local measurements or data 
and uploading to a server in cloud are required.
In cellular IoT,
sensors are to send their local measurements to an access point (AP).
While sensing and uploading can be considered separately,
they can also be integrated, which leads to
data-aided sensing (DAS) \cite{Choi19} \cite{Choi20_IoT}.
In general, DAS can be seen as iterative intelligent data collection scheme
where an AP is to collect data sets from devices or sensor
nodes through multiple rounds.
In DAS, the AP chooses a set of nodes at each round
based on the data sets that are already available at the AP
from previous rounds for efficient data collection.
As a result, 
the AP (actually a server that is connected to the AP)
is able to efficiently provide an answer to a given query 
with a small number of measurements compared to random polling.

In this paper, we apply DAS to distributed
detection in WSNs. Due to the nature of DAS where
sensor nodes sequentially send their measurements
according to a certain order,
distributed detection can be seen as sequential
detection \cite{Wald45}.
Clearly, if sensors' measurements are 
independent and identically distributed (iid),
the performance of distributed detection
is independent of the uploading order.
However, for correlated measurements,
a certain order can allow an early decision
or a reliable decision with a smaller number of sensors'
measurements.
Thus, based on DAS, we can provide 
an efficient way to decide the order of sensor
nodes for uploading, which can 
lead to a reliable decision with a smaller number of 
measurements.
In particular,
we derive an objective function
based on the J-divergence
\cite{Kullback59} 
to decide the next sensor node in DAS
for distributed detection.

The paper is organized as follows. The system model
is presented in Section~\ref{S:SM}.
The notion of DAS is briefly explained and
applied to ditributed detection in Sections~\ref{S:DAS}
and~\ref{S:DD}.
We present simulation results
and conclude the paper with some
remarks in Sections~\ref{S:Sim} and~\ref{S:Con}.

\subsubsection*{Notation}
Matrices and vectors are denoted by upper- and lower-case
boldface letters, respectively.
The superscript $\rT$ denotes the transpose.
$\uE[\cdot]$
stands for the statistical expectation.
In addition, ${\rm Cov}(\bx)$ represents
the covariance matrix of random vector $\bx$.
$\cN(\ba, \bR)$
represents the distributions of
real-valued Gaussian
random vectors with mean vector $\ba$ and
covariance matrix $\bR$, respectively.

\section{System Model} \label{S:SM}

In this section, 
we present the system model for a WSN
that consists of a number of
sensor nodes and one AP that is also an FC for distributed detection
\cite{Varshney97Distributed}.
Thus, throughout the paper, we assume that AP and FC are interchangeable.

\begin{figure}[thb]
\begin{center}
\includegraphics[width=\figwidth]{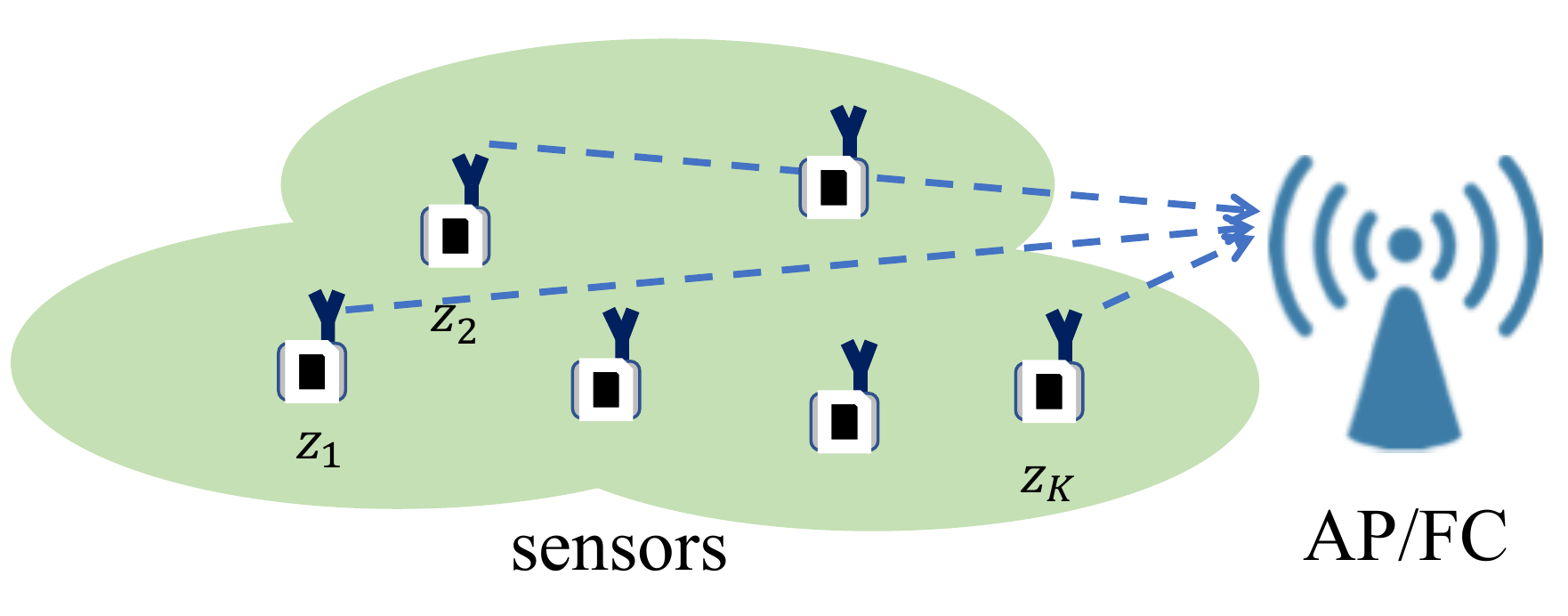}
\end{center}
\caption{A WSN with multiple
sensor nodes and an AP for distributed detection.}
        \label{Fig:wsn}
\end{figure}

Suppose that there are $K$ sensor nodes and 
denote by $z_k$ the local measurement of node $k$
as illustrated in Fig.~\ref{Fig:wsn}.
Each sensor is to send its local measurement to the AP.
The AP is to choose one of $Q$ hypotheses,
which are denoted by $\{\theta_1, \ldots, \theta_{Q}\}$,
based on the received local measurements.
Provided that all $K$ measurements are available,
the AP can make a decision based on the
maximum likelihood (ML) principle as follows:
\begin{align}
q^* = \argmax_q f(z_1, \ldots, z_K \,|\, \theta_q),
	\label{EQ:ML}
\end{align}
where 
$f(z_1, \ldots, z_K \,|\, \theta_q)$ represents
the likelihood function of $\theta_q$ for 
given $\{z_1, \ldots, z_K \}$.

For example, suppose that the $z_k$'s follow
a Gaussian distribution with mean vector $\bm_q$ and 
covariance matrix $\bR_q$ under hypothesis $\theta_q$.
In this case, 
\eqref{EQ:ML} can be re-written as
\begin{align}
q^* & = \argmax_q \ln f(\bz\,|\, \theta_q) \cr
& = \argmin_q (\bz - \bm_q)^\rT \bR_q^{-1}(\bz - \bm_q) 
+ \ln \det (\bR_q),
	\label{EQ:ML2}
\end{align}
where $\bz = [z_1 \ \ldots z_K]^\rT$
that is referred to as the complete information.

For a large number of sensor nodes, $K$,
while a reliable decision
can be made if all the measurements are uploaded to the AP,
the time to receive 
all the measurements can be long due to a limited bandwidth,
which results in a long decision delay.
In particular, if only one sensor node is allowed to transmit
at a slot time (in time division multiple access (TDMA)),
the time to make 
a decision is equivalent to that to receive all the measurements,
i.e., $K$-slot time (under the assumption that
the processing time to choose a hypothesis according
to \eqref{EQ:ML2} is negligible). 
As a result, in order to shorten the decision delay, 
a fraction of measurements can be used at the cost of degraded
decision performance.

\section{Data-aided Sensing}	\label{S:DAS}

In this section, we briefly present the notion of DAS
for joint sensing and communication
to efficiently upload measurements of nodes
\cite{Choi20_IoT} \cite{Choi20_WCNC}.

Suppose that only one
node can upload its measurement at each round. 
Denote by $\cZ (l)$ the set of local measurements
that are available at the AP after iteration $l$. 
Then, for a certain objective function, denoted by $\Theta(\cdot)$,
the AP can decide the next node to upload its local information
as follows:
\be
k (l+1) = \argmin_{k \in \cK^c (l)} \Theta (z_k, \cZ(l)),
	\label{EQ:key}
\ee
where $k(l)$ is the index of the node that is to upload
its local measurement at round $l$,
$\cK(l)$ represents the index set of the nodes associated with
$\cZ(l)$, and $\cK^c (l)$ denotes the complement of $\cK(l)$.
Clearly, it can be shown that
$\cK(l) = \{k(1), \ldots, k (l)\}$.
In \eqref{EQ:key}, we can see that
the selected node, say $k(l+1)$,
at each round depends on
the accumulated measurements
from the previous rounds, $\cZ(l)$, which shows the key idea of DAS.

The objective function in
\eqref{EQ:key} varies depending on the application.
As an example, let us consider an entropy-based
DAS where 
the AP is to choose the uploading order based on the entropy
of measurements.
Since 
$\cZ(l)$ is available at the AP after iteration
$l$, the entropy or information of the remained measurements becomes
$\sH(\cZ^c (l) \,|\, \cZ(l))$,
where $\sH(X|Y)$ represents the conditional entropy of $X$
for given $Y$.
Thus, as in \cite{Choi20_WCNC},
the following objective function can be used for the node selection:
\be
\Theta(z_k, \cZ(l))
=
\underbrace{\sH(\cZ^c (l) \,|\, \cZ(l))}_{\rm Remained \ Information}
- \underbrace{\sH(z_k \,|\, \cZ(l))}_{\rm Updated\ Information},
	\label{EQ:Ent_O}
\ee
which is the entropy gap,
where $\sH(z_k \,|\, \cZ(l))$ is the amount of information 
by uploading the measurement from node $k$.
Clearly, for fast data collection 
(or data collection with a small number of nodes),
we want to choose the next nodes to minimize the entropy gap.

Note that
$\sH(\cZ^c (l) \,|\, \cZ(l))$ is independent of $k$.
Thus, the next node is to be chosen according
to the maximization of conditional entropy is given by
\be
k (l+1) = \argmax_{k \in \cK^c (l)} \sH(z_k\,|\, \cZ(l)).
\ee
That is, the next node should have the maximum
amount of information
(in terms of the conditional entropy)
given that $\cZ(l)$ is already available at the AP.

In \cite{Choi20_IoT},
the mean squared error (MSE) 
between the total measurement $\bz$ 
and its estimate is to be minimized in choosing the node 
at round $l+1$,
i.e.,
\be
\Theta (z_k, \cZ(l))
= \uE \left[||\bz - \hat \bz (z_k, \cZ(l)) ||^2 \, |\, 
\cZ(l) \right],
\ee
where $\hat \bz (z_k, \cZ(l))$ is the minimum MSE 
estimator of $\bz$ for given $\{z_k,\cZ(l)\}$ and 
the expectation is carried out over
$\cZ^c(l)$, which is the subset of measurements
that are not available before round $l$.
Since the measurements are assumed to be Gaussian 
in \cite{Choi20_IoT},
second order statistics can be used for DAS, 
which is called Gaussian DAS.
After some manipulations, it can also be shown that
\be
\Theta (z_k, \cZ(l))
= - \uE \left[ |z_k - \hat z_k |^2 \,\bigl|\, \cZ(l) \right]
+ C,
\ee
where $C$ is constant and $\hat z_k = \uE[z_k \,|\, \cZ(l)]$
is the conditional mean of $z_k$ for given $\cZ(l)$.
Thus, the next node can be selected as follows:
\be
k (l+1) = \argmax_{k \in \cK^c (l)} 
\uE \left[ |z_k - \hat z_k |^2 \,\bigl|\, \cZ(l) \right].
\ee
For Gaussian measurements, 
the MSE,
$\uE \left[ |z_k - \hat z_k |^2 \,\bigl|\, \cZ(l) \right]$,
is proportional to the conditional entropy,
$\sH(z_k \,|\, \cZ(l))$. 
Thus, Gaussian DAS is also to minimize the
entropy gap in selecting nodes, i.e.,
Gaussian DAS is also the entropy-based DAS.

\section{Node Selection in Distributed Detection with DAS}	\label{S:DD}

In this section, we apply DAS to 
distributed detection.
We first consider 
the entropy-based DAS to decide
the next node in distributed detection.
Then, another approach is proposed, which has
the objective function based on the J-divergence.

\subsection{Entropy-based DAS for Node Selection}

Since the entropy-based DAS can
allow the AP to estimate or approximate
the complete information, $\bz$,
with a small number of measurements,
it can be used in distributed detection.
That is, under each hypothesis, 
the AP can collect a subset of measurements
and use them to make a decision.

Suppose that the AP
has a subset of measurements, i.e., $\cZ(l)$, at the end of round $l$.
The AP has the likelihood function,
$f(\cZ(l)\,|\, \theta_q)$.
Under hypothesis $\theta_q$,
the updated likelihood function with a new measurement
from node $k$ becomes
$f(z_k, \cZ(l)\,|\, \theta_q)$,
where $k \in \cK^c (l)$. As in Section~\ref{S:DAS},
for each hypothesis $\theta_q$,
the node selection can be carried out to minimize
the entropy gap.
Thus, if the ascending order is employed,
the node selection
becomes
\be
k (Q i + q) = \argmax_{k \in \cK^c (Qi+q-1)}
\sH(z_k\,|\, \cZ(Qi+q-1), \theta_q),
	\label{EQ:EDAS}
\ee
for $q = 1,\ldots,Q$,
where $Q$ nodes are selected at each $i \in \{0, 1, \ldots\}$. 
Clearly, we have
$ \cK (Q (i+1)) = \cK(Qi) \cup\{k(Qi +1, \ldots, k(Q (i+1))\}$,
and a decision can be made every
$i$, i.e., after $Q$ nodes upload their measurements.

Alternatively, if there are $Q$ parallel channels,
the selected nodes can be given by
\be
k_q (l+1) = \argmax_{k \in \cK^c (l)} \sH(z_k\,|\, \cZ(l), \theta_q),
\ q = 1,\ldots, Q.
\ee

\subsection{KL Divergence for Node Selection}

With known measurements
at the end of round $l$,
i.e., $\cZ(l)$, 
the AP can choose the next node to increase
the log-likelihood ratio (LLR) for the maximum
likelihood (ML) detection. 
For example, suppose that $Q = 2$.
Then, with a new measurement, the LLR is given by
\be
{\rm LLR}_{1,2} (z_k, \cZ(l))  = 
\ln \frac{f (z_k, \cZ(l)\,|\, \theta_1)} 
{f (z_k, \cZ(l)\,|\, \theta_2)} ,
\ k \in \cK^c (l).
\ee
Since $z_k \in \cK^c (l)$
is unknown, the expectation of LLR can be considered.
Under $\theta_1$, the expectation of LLR is given by
\be
\uE[{\rm LLR}_{1,2} (z_k, \cZ(l)) \,|\, \cZ(l), \theta_1],
\ee
and $z_k$ can be chosen to maximize 
the expectation.
However, since there is also hypothesis $\theta_2$, 
for uniform prior,
the objective function to be maximized can be defined as
\begin{align}
\Theta (z_k, \cZ(l))
& = 
\uE[{\rm LLR}_{1,2} (z_k, \cZ(l)) \,|\, \cZ(l), \theta_1] \cr
& \quad +\uE[{\rm LLR}_{2,1} (z_k, \cZ(l)) \,|\, \cZ(l), \theta_2] \cr
& = \sD( f_1 (z_k\,|\, \cZ(l) ) || f_2 (z_k \,|\, \cZ(l) ) ) \cr
& \quad + \sD( f_2 (z_k\,|\, \cZ(l) ) || f_1 (z_k \,|\, \cZ(l) ) ) ,
	\label{EQ:ODD}
\end{align}
where 
$f_q (z_k\,|\, \cZ(l) ) = f(z_k \,|\, \cZ(l), \theta_q)$
and $\sD (f (z) || g (z)) = \uE_f 
\left[\ln \frac{f(Z)}{g(Z)} \right]$ is
the Kullback–Leibler
(KL) divergence between two distributions, $f(z)$ and $g(z)$ 
\cite{Kullback59}.
Here, $\uE_f [\cdot]$ represents
the expectation over $f(z)$.
In \eqref{EQ:ODD},
the objective function
is the $J$-divergence \cite{Kullback59} between
$f_1(z_k\,|\, \cZ(l))$ and $f_2(z_k\,|\, \cZ(l))$.

In general, for $Q$ hypotheses,
the node selection 
can be generalized as follows:
\be
k(l+1) = \argmax_{k \in \cK^c (l)}
\sum_{q=1}^Q \sum_{i \ne q}
\sD (f_q (z_k) || f_i (z_k)).
	\label{EQ:JDAS}
\ee
The resulting DAS that uses
the node selection criterion in \eqref{EQ:JDAS}
is referred to as the J-divergence based DAS (J-DAS).

We now derive more explicit expressions for 
the objective function in \eqref{EQ:ODD} when $\{z_k\}$
are jointly Gaussian.
For simplicity, consider $Q = 2$. Then, the
objective function in \eqref{EQ:ODD} becomes
\begin{align}
\sD(f_1||f_2) + \sD(f_2||f_1)
& = \uE_1 [\ln f_1] - \uE_2 [\ln f_1] \cr
& \quad +  \uE_2 [\ln f_2] - \uE_1 [\ln f_2],
	\label{EQ:DD}
\end{align}
where $f_q = f_q(z_k\,|\, \cZ(l))$
and $\uE_q [\cdot] = \uE_{f_q} [\cdot]$
for notational convenience.

For convenience, let $\bz(l)$ be the column
vector
consisting of the elements of $\cZ(l)$.
In addition, let
$\bar z_{k;q} = \uE_q [z_k]$,
$\bar \bz_q (l) = \uE_q [\bz(l)]$,
and $\bR_q (l) = \uE_q [
\left(\bz(l) - \bar \bz_q(l) \right)
\left(\bz(l) - \bar \bz_q(l) \right)^\rT ]$.
Let $r_{k;q} = \uE_q [(z_k - \bar z_{k;q})^2]$ and
$\bc_{k;q} (l) =
\uE_q \left[
\left(\bz(l) - \bar \bz_q(l) \right) (z_k - \bar z_{k;q}) \right]$,
where $\bar z_{k;q} = \uE_q [z_k]$.
Then, under $\theta_q$, we have
\be
\left[
\begin{array}{c}
z_k \cr
\bz(l) \cr
\end{array}
\right]
\sim \cN \left(
\left[
\begin{array}{c}
\bar z_{k;q} \cr
\bar \bz_q(l) \cr
\end{array}
\right],
\left[
\begin{array}{cc}
r_{k;q} & \bc_{k;q}^\rT (l) \cr
\bc_{k;q} (l) & \bR_q (l) \cr
\end{array}
\right]
\right).
\ee

After some manipulations using
the matrix inversion lemma \cite{Harv97}, it can be shown that
\begin{align}
\uE_1 [\ln f_1] -\uE_2 [\ln f_1] 
& = -\frac{\alpha_{k;1}}{2} \left(
r_{k;1} - r_{k;2} - \bar d_k^2
\right) \cr
& \quad - \bar d_k (\bz (l) - \bar \bz_1 (l))^\rT \bb_{k; 1} ,
	\label{EQ:EEf}
\end{align} 
where
$\bar d_k = \bar z_{k;1} - \bar z_{k;2}$ and
\begin{align}
\alpha_{k;q} 
& = \frac{1}{r_{k;1} -  \bc_{k;q}^\rT (l) \bR_q^{-1} (l) \bc_{k;q} (l)}\cr
\bb_{k;q} & = -\alpha_{k;q} \bR_q^{-1} (l) \bc_{k;q} (l) .
	\label{EQ:ab}
\end{align}

The second term in \eqref{EQ:DD}, i.e.,
$\uE_2 [\ln f_2] - \uE_1 [\ln f_2]$, can also be found by
a similar way.

For Gaussian measurements, the complexity of J-DAS
mainly depends on the complexity to 
find the J-divergence or \eqref{EQ:EEf}.
At round $l+1$, 
since $\bz (l) - \bar \bz_1 (l)$ is the common term
for all $k \in \cK^c (l)$, the complexity is mainly due to
that to find $\bb_{k;q}$ in \eqref{EQ:ab}.
Again, $\bR_q (l)$ is independent of $k$,
the complexity to find $\bb_{k;q}$ is $O(l^2)$, 
while the complexity to find the inverse of $\bR_q (l)$
from the results in the previous round is $O(l^2)$.
Thus,
the complexity at round $l+1$ is $O((K-l)l^2)$.
Thus, if the number of iterations is $T (\ll K)$, the total
complexity is bounded by $O( K \sum_{l=1}^{T} l^2) = O(K T^3)$.

\section{Simulation Results}	\label{S:Sim}

In this section, we present simulation results
with LLRs. For each average LLR,
the results of 500 independent runs are used.
In simulations, we assume that
the $z_k$'s are jointly Gaussian with $K = 50$
and $Q = 2$. For comparisons,
we consider the three node selection
criteria: J-DAS
(i.e., \eqref{EQ:JDAS}), 
entropy-based DAS
(i.e., \eqref{EQ:EDAS}), and random selection.

\begin{figure}[thb]
\begin{center}
\includegraphics[width=\figwidth]{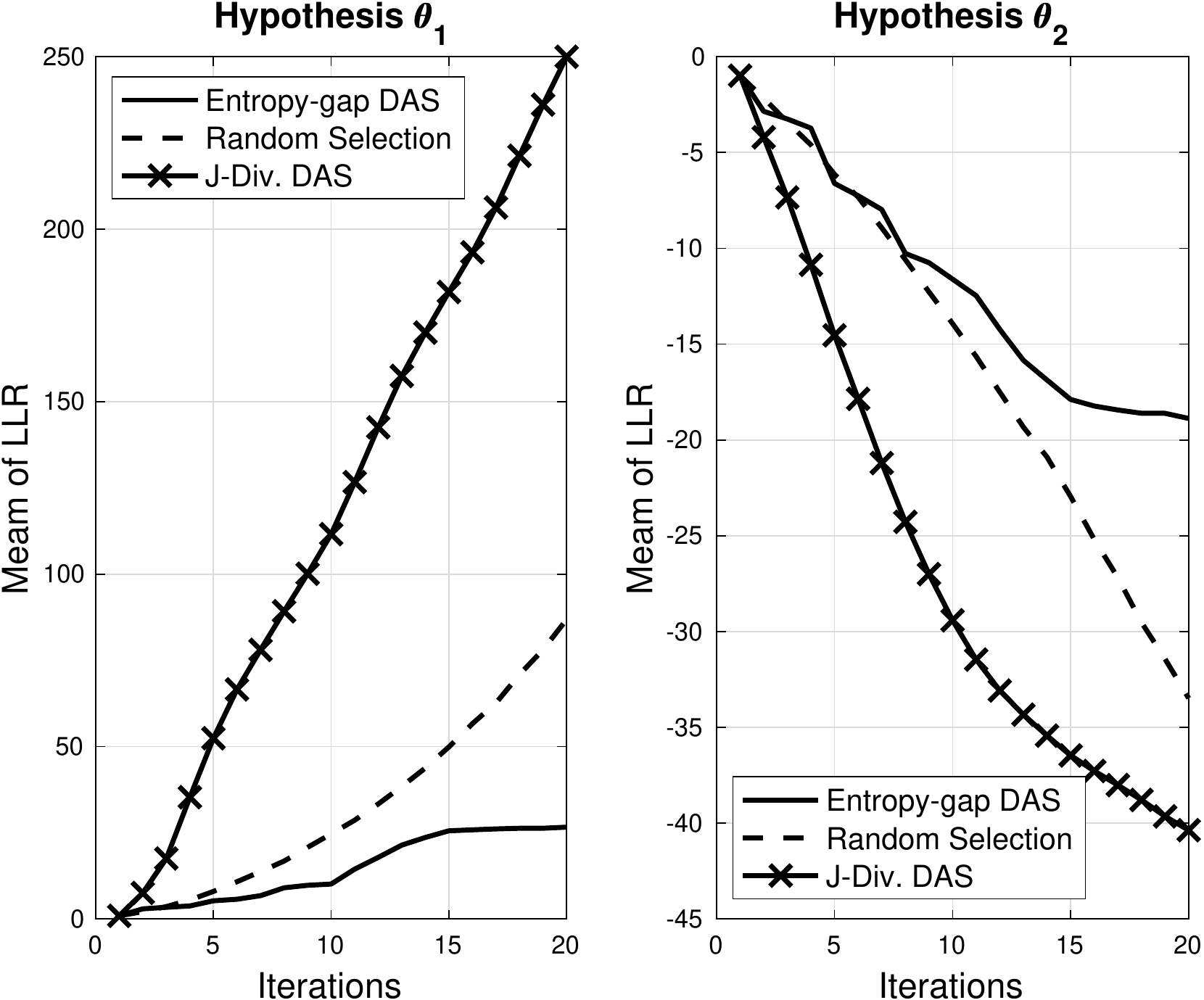} \\
\hskip 0.5cm (a) \hskip 3.5cm (b)
\end{center}
\caption{The sample mean of LLR 
when the $z_k$'s are jointly Gaussian 
when the means and covariance matrices
are given by \eqref{EQ:M1} and \eqref{EQ:C1}, respectively, with
$\sSNR = 0$ dB and $K = 50$:
(a) when $\theta_1$ is true; (b) when $\theta_2$ is true.} 
        \label{Fig:plt1}
\end{figure}

For the first simulation,
we assume that
\begin{align}
\bar z_{k;1} = A \cos \left( \frac{\pi}{10}(k- 1) \right), \ 
\bar z_{k;2} = A \sin \left( \frac{\pi}{10}(k- 1) \right),
	\label{EQ:M1}
\end{align}
where $A > 0$ is the amplitude of the mean signal.
In addition, the covariance matrix of $\bz$ under $\theta_q$ is given by
\be
[\bR_q]_{k,t} = \sigma^2 (\rho_q)^{|k-t|},
\ k,t = 1, \ldots, K,
	\label{EQ:C1}
\ee
where $\rho_1 = \frac{3}{4}$ and $\rho_2 = - \frac{3}{4}$,
and $\sigma^2$ represents the noise variance.
The signal-to-noise ratio (SNR)
is defined as $\sSNR = \frac{A^2}{2 \sigma^2}$.
In Fig.~\ref{Fig:plt1},
the average LLR is shown as a function of iterations.
When $\theta_1$ is correct,
it is shown that the LLR increases with the number of samples
or iterations.
Clearly, the performance of J-DAS outperforms 
those of entropy-based DAS and random selection.
In other words, when a decision is made with a small
number of iterations or samples in distributed detection,
J-DAS can help provide reliable outcomes
compared to the others.

It is interesting to see that the entropy-based DAS
does not provide a good performance
in distributed detection as shown in Fig.~\ref{Fig:plt1}.
The entropy-based DAS
is to choose the node that has the most uncorrelated
measurement as shown in \eqref{EQ:EDAS}.
In fact, such a measurement 
needs not contribute to maximizing the difference between the hypotheses.

\begin{figure}[thb]
\begin{center}
\includegraphics[width=\figwidth]{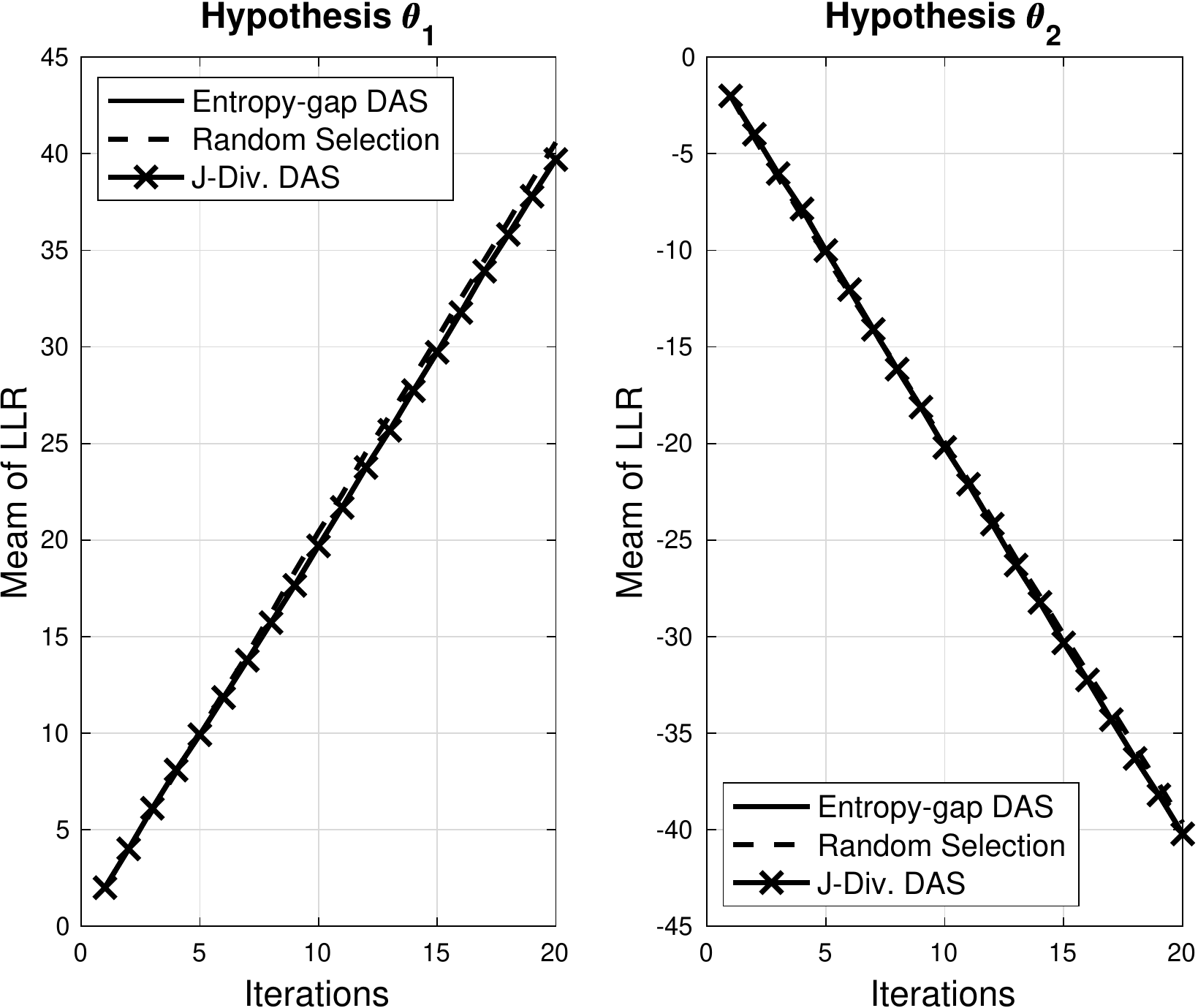} \\
\hskip 0.5cm (a) \hskip 3.5cm (b)
\end{center}
\caption{The sample mean of LLR 
when the $z_k$'s are iid Gaussian with
$\sSNR = 0$ dB and $K = 50$:
(a) when $\theta_1$ is true; (b) when $\theta_2$ is true.} 
        \label{Fig:plt3}
\end{figure}

For the next simulation,
we consider the case of iid measurements.
That is, $\bR_q = \sigma^2 \bI$ for all $q \in \{1,2\}$.
In addition, it is assumed that
$\bar z_{k;1} = A$ and $\bar z_{k;2} = -A$ for all $k$.
As mentioned earlier,
in this case, the performance is independent of
the uploading order,
which can be confirmed by the results in
Fig.~\ref{Fig:plt3}.

\section{Concluding Remarks}	\label{S:Con}

In this paper, the notion of DAS has been exploited
to provide a good performance of distributed
detection in WSN with a smaller number of 
sensors that upload their measurements.
In particular, a J-divergence based node selection
criterion was proposed for DAS, which was referred to as J-DAS,
so that the measurement from the next node
can maximize LLR.
From simulation results, 
it was shown that LLR increased rapidly 
as the number of iterations increased 
when J-DAS was employed.

There would be a number of further research topics. Among those,
as in \cite{Choi20_WCL}, the application of J-DAS 
to distributed machine learning
for classification is interesting. The performance
analysis of J-DAS would also be another topic
to be studied in the future.

\bibliographystyle{ieeetr}
\bibliography{sensor}

\end{document}